%% file: main.tex
\documentclass{iopart}
\usepackage{iopams}
\usepackage{citesort}
\usepackage[dvips]{graphicx}
\usepackage{xcolor}
\usepackage{mathrsfs}
\usepackage{subeqn}
\usepackage[fleqn]{cases}
\usepackage{float}

\newcommand{\eqind}{\,{\buildrel d \over =}\,}
\newcommand{\binom}[2]{{ #1 \choose #2}}
\newcommand{\dif}[2]{{\rmd{#1} \over \rmd{#2}}}

\begin{document}
\title[]{Fokker-Planck Equation and Path Integral Representation of Fractional Ornstein-Uhlenbeck Process with Two Indices}

\author{Chai Hok Eab}
\address{%
Department of Chemistry
Faculty of Science, Chulalongkorn University
Bangkok 10330, Thailand%
}
\ead{Chaihok.E@chula.ac.th}

\author{S.C. Lim}
\address{%
Faculty of Engineering, Multimedia University
 63100 Cyberjaya, Selangor Darul Ehsan, Malaysia%
}
\ead{sclim47@gmail.com}

\begin{abstract}
This paper considers the Fokker-Planck equation and path integral formulation of the fractional Ornstein-Uhlenbeck process parametrized by two indices. 
The effective 
Fokker-Planck equation of this process is derived from the associated fractional Langevin equation.
Path integral representation of the process is constructed and the basic quantities are evaluated.
\end{abstract}

\pacs{02.50.Ey, 05.10.Gg, 05.40.-a}

\maketitle

\input{introduction}

\input{fracOU2}

\input{pathOU2}

\input{conclude}


\appendix
\input{FPeqPotential}

\input{pathOU2detail}

\input{CaputoRL}

\section*{References}
\bibliographystyle{iopart-num}
\bibliography{biblioPathOU2}

\end{document}

%% file: introduction.tex
\section{Introduction}
\label{sec:introduction}
The two Gaussian Markov processes, Brownian motion and Ornstein-Uhlenbeck process, have been used extensively in various applications from natural sciences to financial mathematics. 
However, many systems in the real world are more complex and they are non-Markovian in character with memory.  Therefore it is necessary to go beyond these simple Markovian models based on Brownian motion and Ornstein-Uhlenbeck process. 
For example, Brownian motion as the model for normal diffusion process can be generalized to fractional Brownian motion 
\cite{Biagini08,Mishura08,Nourdin12}
in order to describe anomalous diffusion 
\cite{Kalages08}.
Similarly, models in financial time series and meteorology  
based on ordinary Ornstein-Uhlenbeck process 
has to use its Gaussian fractional generalization, the fractional Ornstein-Uhlenbeck process 
\cite{Cheridito03,LimMuniandy03,Magdziarz08,YanLuXu08}.

In many applications of Brownian motion and Ornstein-Uhlenbeck process, path integral method has played an important role 
\cite{Chaichian01,Kleinert09,Wio13a}.
One would expect path integral technique to have a similar role for fractional Brownian motion and fractional Ornstein-Uhlenbeck process. 
Lately, there has been considerable interest in the applications of 
path integral method
in quantum mechanics and quantum field theory in fractal and multifractal spacetime. 
Applications of path integral in fractional quantum mechanics was studied by Laskin 
\cite{Laskin00}
and several authors  
\cite{TarasovZaslavsky08,CalcagniNardelliScalisi12,KleinertZatloukal13}.
Path integral formulation has also been used in fractional quantum field theory 
\cite{Kleinert12,Ferrante13}.
Connection between fractional stochastic calculus and constructive field theory has also been considered 
\cite{MagnenUnterberger11,MagnenUnterberger12}.
Another motivation of studying fractional and multifractional path integrals is that several candidate theories of quantum gravity share the idea that spacetime is multifractal with integer dimension 4 at large scales, while it is two-dimensional in the ultraviolet limit or small scales
\cite{Modesto09,Ambjorn10,Calcagni10}.
This leads to the necessity to consider quantum theory in multifractal spacetime 
\cite{Calcagni13,ReuterSaueressig12,Calcagni12}.

Due to the presence of fractional
integro-differential operators,
one would expect the evaluation of the path integrals of the fractional stochastic processes to be more complicated. 
Sebastian \cite{Sebastian95} was first to study the path integral representation of fractional Brownian motion. 
Subsequently path integral of fractional Brownian motion has been applied to model polymers 
\cite{CherayilBiswas93,ChakravartiSebastian97}.
Lately, there has been renewed interest in path integral formulation of fractional Brownian motion and fractional Levy process by several authors
\cite{CalvoSanchez08,CalvoSanchezCarreras09,JanakiramanSebastian12,Wio13b}.
Path integral for fractional Ornstein-Uhlenbeck process or fractional oscillator process with single index has been considered in 
\cite{EabLim06}.
Another approach of path integral
has been considered by Friedrich and Eule who used discrete time path integral representation for the continuous time random walk 
\cite{FriedrichEule11,EuleFriedrich07}.
The Eule-Friedrich  approach is quite different from the one used in our work,
which follows that of Sebastian \cite{Sebastian95}.
A brief summary of the path integral representations of various fractional processes is given in the book 
\cite{Wio13a}.

In the application to systems which have variable memory it is necessary to use a fractional process parametrized by an index which varies in time or space, for example, the multifractional Brownian motion 
\cite{PeltierVehel95,BenassiJaffardRoux97} 
and multifractional Ornstein-Uhlenbeck process 
\cite{LimTeo07}.
However, such multifractional processes are more complex and less adapted to practical purposes. It may be good to have a process that can provide a more flexible model and yet is mathematically more tractable. 
In the case of Ornstein-Uhlenbeck process, instead of multifractional Ornstein-Uhlenbeck process one can use fractional Ornstein-Uhlenbeck process indexed by two parameters. 
The additional index for fractional Ornstein-Uhlenbeck process allows more flexibility in applications. 
For the fractional Ornstein-Uhlenbeck process with single index,
both its long-time and short-time properties are characterised 
by a single parameter. The main
advantage of fractional Ornstein-Uhlenbeck process with two indices over that of single index is
that both its long-time and short-time behavior can have separate characterization by two
different parameters. Possible applications of fractional Ornstein-Uhlenbeck process with two
indices include the modeling of the Von Karman wind speed spectrum 
\cite{LimLiTeo08}
and the study of Casimir energy for fractional quantum field 
\cite{LimTeo09,LimTeo11}.

In this paper we first consider briefly the properties of fractional Ornstein-Uhlenbeck process with two indices, which will be followed by the derivation of the Fokker-Planck equation for the process. 
Subsequent section contains the path integral formulation for the fractional Ornstein-Uhlenbeck process with two indices.


%% file: fracOU2.tex
\section{Fractional Ornstein-Uhlenbeck process with two indices}
\label{sec:fracOU2}
First we recall that ordinary Ornstein-Uhlenbeck process can be obtained as 
the solution of the usual Langevin equation
\begin{equation}
  D_tx(t) + \lambda x(t)  = \xi(t), 
\label{eq:fracOU2_0010}
\end{equation}
where $\xi(t)$  is standard white noise with its mean zero and covariance given by 
$\bigl<\xi(t)\xi(s)\bigr>=\delta(t - s)$. 
Solution of (\ref{eq:fracOU2_0010}) is 
\begin{equation}
  x(t)=x_\circ e^{-\lambda{t}} + \int_0^t G(t-u)\xi(u)du,
\label{eq:fracOU2_0020}
\end{equation}
where $x_\circ = x(0)$ and $G(t)=e^{-\lambda{t}}$. 
One gets
\begin{eqnarray}
  \bigl<x(t)\bigr> & = & x_\circ e^{-\lambda{t}}, 
\label{eq:fracOU2_0030}
\\
  \bigl<x(t)x(s)\bigr> & = & \frac{1}{2\lambda}\Bigl(e^{-\lambda(t-s)} - e^{-\lambda(t+s)}\Bigr) .
\label{eq:fracOU2_0040}
\end{eqnarray}
In the long-time limit, the Ornstein-Uhlenbeck process becomes a stationary process with covariance
\begin{equation}
  \bigl<x(t)x(s)\bigr>  = \frac{e^{-\lambda(t-s)}}{2\lambda} .
\label{eq:fracOU2_0050}
\end{equation}
In other words, the Ornstein-Uhlenbeck process is a stationary process if the starting time 
$t =-\infty$; and a non-stationary process if it begins at a finite time.
Such a dependence on initial time of the time integro-differential operators is reflected in the distinction between Riemann-Liouville (and also Caputo) fractional derivative and the Weyl fractional derivative 
\cite{LimEab06}.

The Fokker-Planck equation corresponds to Ornstein-Uhlenbeck process is given by
\begin{equation}
  \frac{\partial}{\partial{t}} P\bigl(x,t|x_\circ, t_\circ\bigr) = \frac{\partial}{\partial{x}} \Big[\lambda{x}P\bigl(x,t|x_\circ, t_\circ\bigr)\Bigr] + \frac{D}{2}\frac{\partial^2}{\partial{x}^2} P\bigl(x,t|x_\circ, t_\circ\bigr),
\label{eq:fracOU2_0060}
\end{equation}
which corresponds to the Lagrangian
\label{pl:fracOU2_0010}
\begin{equation}
  L\bigl(x,\dot{x}\bigr) = \frac{1}{2}\bigl(\dot{x}+ \lambda{x}\bigr)^2,
\label{eq:fracOU2_0070}
\end{equation}
where $\dot{x}$ denotes ordinary time derivative of $x$. 
The Euler-Langrange equation is
\begin{equation}
  \ddot{x} - \lambda^2{x} = 0 .
\label{eq:fracOU2_0080}
\end{equation}

The path integral representation of the Ornstein-Uhlenbeck process, which is a
Markov process can be obtained quite easily just like the case for Brownian motion
\cite{Ezawa00}.  
By using the Kolmogorov equation, the infinitesimal propagator can be chained into a path integration.
One gets
\begin{equation}
\fl\quad
G_\circ\bigl(x_N,t_N|x_\circ,t_\circ\bigr) = \int \cdots \int 
                                               \left\{
                                                   \prod_{k=1}^{N-1} G_\circ\bigl(x_{k+1},t_{k+1}|x_k,t_k\bigr)dx_k
                                                 \right\}
                                                   G_\circ\bigl(x_1,t_1|x_\circ,t_\circ\bigr),
 \label{eq:EzawaEqbefor217} 
\end{equation}
where
\begin{equation}
\fl \qquad\quad
  G_\circ\bigl(x,t|x_\circ,t_\circ\bigr) = \frac{1}{\sqrt{2\pi\lambda{D}\bigl(1 - e^{2\lambda{(t-t_\circ)}}\bigr)}}
                                   \exp\left(-\frac{\bigl(x - x_\circ e^{-\lambda{(t-t_\circ)}}\bigr)^2}{2\lambda{D}\bigl(1 - e^{2\lambda{(t-t_\circ)}}\bigr)}\right).
\label{eq:fracOU2_0090}
\end{equation}
By taking limit $N \to \infty$ and $|t_{k+1} - t_k| \to 0$ while keeping $\sum_{k=0}^{N-1} |t_{k+1} - t_k| < \infty$ in {(\ref{eq:EzawaEqbefor217})},
one gets the path integration of the Ornstein-Uhlenbeck process 
\cite{Ezawa00}.

There are several ways of generalize Ornstein-Uhlenbeck process to its fractional counterpart. 
One way is to replace the white noise by a fractional Gaussian noise 
\cite{Cheridito03,YanLuXu08},
or one can apply the Lamperti transformation to fractional Brownian motion 
\cite{LimMuniandy03,Magdziarz08}.
Note that if the white noise in the Langevin equation is replaced by the fractional Levy motion, its solution is fractional Ornstein-Uhlenbeck process of $\alpha$-stable-type 
\cite{BelBarkai05,DengBarkai09}.
In this paper we shall not consider all these variants of fractional Ornstein-Uhlenbeck processes. Instead, we shall restrict to the fractional Ornstein-Uhlenbeck process obtained from the fractional Langevin equation by keeping
the white noise and replace the time-differential operator 
$D_t$ by a fractional one $D_t^\alpha$ in the Langevin equation 
(\ref{eq:fracOU2_0010}).
There exist the following three possible fractional generalizations of Langevin equation:
\begin{subequations} 
\label{eq:fracOU2_0100}
  \begin{equation}
    D_t^\alpha x(t) + \lambda^\alpha x(t) = \xi(t) ,
    \label{eq:fracOU2_0100a}
  \end{equation}
  \begin{equation}
    \bigl(D_t + \lambda\bigr)^\gamma x(t) = \xi(t) ,
    \label{eq:fracOU2_0100b}
  \end{equation}
and
  \begin{equation}
    \bigl(D_t^\alpha + \lambda^\alpha\bigr)^\gamma x(t) = \xi(t) ,
    \label{eq:fracOU2_0100c}
  \end{equation}
\end{subequations}
where $0 < \alpha \leq 1$ and $0 < \gamma \leq 1$.
Note that one can formally defined the ``shifted'' fractional derivative
$\bigl(D_t^\alpha + \lambda^\alpha\bigr)^\gamma$
in terms of the unshifted derivative
$D_t^\alpha$.
By using binomial expansion, it is possible
to express the shifted fractional derivative in terms of unshifted ones:
\begin{equation}
\bigl(D_t^\alpha + \lambda^\alpha\bigr)^\gamma = \sum_{j=0}^\infty \binom{\gamma}{j}
                                              \lambda^{\alpha{j}} D_t^{\alpha(\gamma-j} ,
\label{eq:revise_fracOU2_0101}
\end{equation}
with the fractional derivative of the Caputo and Riemann-Liouville type defined 
as follows.
The fractional derivative of order $\alpha$ denoted by ${_a}D_t^\alpha$ 
can be defined in terms of its inverse operator or the fractional integral 
\cite{SamkoKibasMarichev93}:
\begin{equation}
  \label{eq:revisefracOU2_0102}
 {_a}D_t^{-\alpha}f(t) = {_a}I_t^\alpha f(t) = \frac{1}{\Gamma(\alpha)}\int_a^t (t-u)^{\alpha-1} f(u)du.
\end{equation}
When $a = -\infty$ the fractional derivative is known as Weyl fractional derivative; when $a = 0$ 
(for simplicity we write ${_0}D_t^\alpha \equiv D_t^\alpha$), one gets the Riemann-Liouville and Caputo fractional derivative
according to the following definitions.
A fractional derivative of arbitrary order $\alpha$, with $n-1 \leq \alpha <n$,
can be defined through fractional integration
of order $n-\alpha$ and successive ordinary derivative of order n:
\begin{subequations}
\begin{eqnarray}
  D_t^\alpha f(t) = \left(\dif{}{t}\right)^n D_t^{\alpha-n}f(t),  \qquad \textrm{Riemann-Liouville}
  \label{eq:revisefracOU2_0103a}  \\
  {^C}D_t^\alpha f(t) =  D_t^{\alpha-n}\left(\dif{}{t}\right)^nf(t), \qquad \textrm{Caputo}.
  \label{eq:revisefracOU2_0103b}
\end{eqnarray}  
\end{subequations}

The Lagrangian associated with the Ornstein-Uhlenbeck process is given by
\begin{subequations}
\label{eq:fracOU2_0110}
\begin{equation}
  L = x(t)\Lambda_{\alpha\gamma}\bigl(D_t\bigr)x(t) ,
\label{eq:fracOU2_0110a}
\end{equation}
with
\begin{equation}
   \Lambda_{\alpha\gamma}\bigl(D_t\bigr) = 
        \Bigl(\bigl(D_t^{\alpha}\bigr)^\dagger + \lambda^{\alpha}\Bigr)^\gamma 
        \Bigl(D_t^{\alpha} + \lambda^{\alpha}\Bigr)^\gamma  .
\label{eq:fracOU2_0110b}
\end{equation}
\end{subequations}
Here we remark that a more rigorous treatment of fractional operators such as
$\bigl(D_t^\alpha + \lambda^\alpha\bigr)^\gamma$
and $\Lambda_{\alpha\gamma}\bigl(D_t\bigr)$
can be obtained by using hypersingular integrals 
\cite{Samko02}. 
Path integral representation of fractional Ornstein-Uhlenbeck process with single index given by (\ref{eq:fracOU2_0100b}) has been considered 
\cite{EabLim06}.
In this paper we intend to study the
general case 
(\ref{eq:fracOU2_0100c})
with $\alpha \neq 1$  and $\gamma \neq 1$. We shall restrict to the case of Riemann-
Liouville type of fractional oscillator as it is usually done for path integral formulation of
fractional processes.

\subsection{Basic Properties of Fractional Ornstein-Uhlenbeck Process}
\label{sec:BasicfracOU}
Since $x_{\alpha\gamma}(t)$ is a Guassian process, it can be characterized by its mean and covariance.
For convenience the process is assumed to be centred with zero mean. 
The covariance of the fractional Ornstein-Uhlenbeck process with two indices based on
Riemann-Liouville fractional derivatives is given by 
\cite{LimLiTeo08,LimTeo09}:
\begin{eqnarray}\fl
  C_{\alpha\gamma}(t,s)  =  \Bigl<x_{\alpha\gamma}(t)x_{\alpha\gamma}(s)\Bigr>  \nonumber \\
  \fl \qquad\quad \ \                     =  \sum_{m=0}^\infty\sum_{n=0}^\infty 
                           \binom{\gamma+m-1}{m}\binom{\gamma+n-1}{n}
                           \bigl(-\lambda^\alpha\bigr)^{m+n}                      
                           \frac{s^{\alpha(\gamma+m)}t^{\alpha(\gamma+n)-1}}
                                {\Gamma\bigl(1+\alpha(\gamma+m)\Gamma\bigl(\alpha(\gamma+n)\bigr)}
                                \nonumber \\
                                \fl \qquad \qquad \qquad
                                {{}_2F_1}\Bigl(1 - \alpha(\gamma+n),
                                1, 1 + \alpha(\gamma +m);
                                \frac{s}{t}\Bigr) .
\label{eq:BasicfracOU_0010}
\end{eqnarray}
The variance of $x_{\alpha\gamma}$ is
\begin{eqnarray}
  \sigma^2_{\alpha\gamma}(t) & = & 
  \Bigl<\bigl(x_{\alpha\gamma}(t)\bigr)^2\Bigr> \nonumber \\
  & = & \sum_{q=0}^\infty \bigl(-\lambda^\alpha\bigr)^q
  \frac{t^{\alpha(2\gamma+q)-1}}{\alpha(2\gamma + q) - 1} \Lambda_q ,
\label{eq:BasicfracOU_0020}
\end{eqnarray}
where
\begin{equation}
\fl \qquad\qquad
  \Lambda_q = \sum_{m+n=q} 
                           \binom{\gamma+m-1}{m}\binom{\gamma+n-1}{n}
                           \frac{1}
                                {\Gamma\bigl(\alpha(\gamma+m)\Gamma\bigl(\alpha(\gamma+n)\bigr)} .
\label{eq:BasicfracOU_0030}
\end{equation}

Before we proceed to derive the path integral representation of
$x_{\alpha{\gamma{}}} (t)$ we consider briefly some basic properties
of this process
\cite{LimLiTeo08,LimTeo09}.
First we note that just like all the
Riemann-Liouville processes (such as the case of fractional Brownian
motion
\cite{Lim01} 
and fractional Ornstein-Uhlenbeck process with single index
\cite{LimEab06}),
$x_{\alpha{\gamma{}}} (t)$ is a non-stationary
process and it does not have stationary increments.

Next, we consider the link between fractional Ornstein-Uhlenbeck process and
fractional Brownian motion. Let the increment process 
$x_{\alpha\gamma} (t+\tau) - x_{\alpha\gamma}(t)$
be denoted by
$\Delta x_{\alpha\gamma}(t,\tau)$.
Direct computation one shows that as $\tau \to 0$
the covariance of
$\Delta x_{\alpha\gamma}(t,\tau)$
approaches that of fractional Brownian motion of Hurst index
$H = \alpha\gamma -1/2$. 
That is, as $\tau_1 \to 0$ and $\tau_2 \to 0$,
\begin{equation}
\fl\qquad\quad
  \bigl<\Delta{x}_{\alpha\gamma}(t,\tau_1)\Delta{x}_{\alpha\gamma}(t,\tau_2)\bigr> \sim |\tau_1|^{2\alpha\gamma-1} + |\tau_2|^{2\alpha\gamma-1}
                                                                                  - |\tau_1 - \tau_2|^{2\alpha\gamma-1} .
\end{equation}
Recall that a stochastic process $w(t)$ is locally asymptotically
self-similar 
with index $\kappa$
at the point $t_\circ$ if there exists a non-degenerate process 
$T_{t_\circ}(u)$ such that
\begin{equation}
  \lim_{\epsilon \to 0} \frac{w(t_\circ + \epsilon{u}) -
    w(t_\circ)}{\epsilon^\kappa} \eqind T_{t_\circ}(u) ,
\label{eq:BasicfracOU_0040}
\end{equation}
where $\eqind$ denotes equality of finite dimensional distributions.
Again, by direct computation of the covariance of the increment process on 
the l.h.s.\@ of  
(\ref{eq:BasicfracOU_0040}), 
one can verify that
$x_{\alpha{\gamma{}}} (t)$ is locally
asymptotically self-similar and its tangent process $T_{t_\circ} (u)$ 
is a fractional Brownian motion with Hurst index 
$H = \kappa = \alpha\gamma -1/2$.
It can also be verified that the fractal dimension of 
$x_{\alpha\gamma}(t)$ is 
$5/2 - \alpha\gamma$,
since it locally behaves like fractional Brownian motion which has a fractal dimension of
$2-H$.

Both Brownian motion and Ornstein-Uhlenbeck process are Markov processes. When Brownian motion is generalized to fractional Brownian motion, 
it loses the Markovian property and becomes a long memory (or long-range dependent) process. 
On the other hand, the particular type of fractional extension of Ornstein-Uhlenbeck process considered here is a non-Markov process with short memory (short-range dependence)
\cite{LimEab06}.
There exists another type of fractional Ornstein-Uhlenbeck process which has long range dependent property
\cite{Cheridito03,Magdziarz08,YanLuXu08}. 
See reference
\cite{Magdziarz08}
for other definitions of fractional Ornstein-Uhlenbeck processes and their properties.
Here we would like to add that if the Langevin equation is driven by fractional levy motion, 
its solution will give fractional Ornstein-Uhlenbeck process of $\alpha$-stable-type, 
which is non-Gaussian and its memory structure requires somewhat different characterization 
\cite{MagdziarzWeron07,Magdziarz09}.

Another interesting property is the ergodic property which has recently attracted
considerable attention. For examples, the validity or violation of ergodic property has been
considered for various stochastic processes such as continuous time random walk 
\cite{BelBarkai05},
fractional Brownian motion 
\cite{DengBarkai09}
and fractional Ornstein-Uhlenbeck process $\alpha$-stable-type 
\cite{Magdziarz09}.
Ergodicity is a fundamental property of dynamical system, and its underlying idea is that for a system, the
ensemble average of its properties equals the time average.

Here, we shall verify the ergodicity of the fractional Ornstein-Uhlenbeck process with two
indices using the Khinchin theorem on the condition on its auto-correlation function 
\cite{MagdziarzWeron11,Khinchin48}.
The autocorrelation
function of a centred Gaussian process $y(t)$ is given by
\begin{equation}
  r(s,t) = \frac{\bigl<y(s)y(t)\bigr>}{\Bigl(\bigl<y(s)^2\bigr>\bigl<y(t)^2\bigr>\Bigr)^{1/2}}.
\label{eq:fracOU2ergodic0010}
\end{equation}
According to Khinchin Theorem,  
if the autocorrelation function of a stationary Gaussian $y(t)$ satisfies
$\lim_{t \to \infty} r(t) = 0$,
then $y(t)$ is ergodic.
Fractional Ornstein-Uhlenbeck process of Riemann-Liouville type
$x_{\alpha\gamma}(t)$
is non-stationary.
However, it has been shown that if
$x_{\alpha\gamma}(t)$
is allowed to evolve for sufficient long time, 
it will achieve stationarity and become fractional Ornstein-Uhlenbeck process of Weyl type which is stationary
\cite{LimLiTeo08}.
Since the condition for ergodicity on autocorrelation is in the large-time limit, 
we can verify the ergodic property for the Weyl-type fractional Ornstein-Uhlenbeck process to infer that same property also
holds for 
$x_{\alpha\gamma}(t)$.
It has been shown in ref.\ \cite{LimLiTeo08} that the large time behavior of the covariance
$C(t,t+\tau)$ of $x_{\alpha\gamma}(t)$ varies as $\tau^{-\alpha-1}$, which tends to $0$ as $\tau \to \infty$.
This property, together with with
the finite variance of the process, one verifies that the autocorrelation of 
$x_{\alpha\gamma}(t)$
satisfies the condition $\lim_{t\to\infty}r(t) = 0$ and the process is ergodic 
\cite{MagdziarzWeron11,Khinchin48}.
Thus we see just like in the case of fractional Brownian motion, fractional Ornstein-Uhlenbeck process with two indices also satisfies the ergodic property.

\subsection{Fokker-Planck equation of fractional Ornstein-Uhlenbeck process}
\label{sec:FPeqfracOU}
In this section we would like to derive the Fokker-Planck equation for the fractional Ornstein-Uhlenbeck process $x_{\alpha\gamma} (t)$ from the fractional Langevin equation 
(\ref{eq:fracOU2_0100c})
or any equations derived from (\ref{eq:fracOU2_0100c}).
By adapting an argument similar to that given in reference 
\cite{Reichl98}
(see page 268, Chapter 5, S5.C.3.),
which assumes
strong frictional limit with large frictional coefficient such that the particle relaxes to its stationary state very rapidly.
Consequently, one can assume velocity $v$ does not vary with time such that 
$\dif{v}{t} \approx 0$
and one can then write the corresponding Langevin equation as
\begin{equation}
  \mathbf{K}_t x = F(x) + \xi,
\label{eq:FPeqfracOU_0010}
\end{equation}
where $\mathbf{K}_t$ is a differential-integral operator. 
For Brownian particle, 
fractional Brownian motion and fractional Ornstein-Uhlenbeck process with two
indices $\mathbf{K}_t$ is respectively given by 
(\ref{eq:FPeqfracOU_0020a})-(\ref{eq:FPeqfracOU_0020c})
if $\xi$ is white noise:
\begin{subequations}
\label{eq:FPeqfracOU_0020}
\begin{numcases}
  {\mathbf{K}_t = }     D_t                                      \label{eq:FPeqfracOU_0020a}\\
                        D_t^{H+1/2}                               \label{eq:FPeqfracOU_0020b}\\
                        \bigl(D_t^\alpha + \lambda^\alpha\bigr)^\gamma .  \label{eq:FPeqfracOU_0020c}
\end{numcases}
\end{subequations}
(\ref{eq:FPeqfracOU_0010})
can also be written as
\begin{equation}
  D_t x = D_t \mathbf{L} F(x) + D_t \mathbf{L} \xi ,
\label{eq:FPeqfracOU_0030}
\end{equation}
where $\mathbf{L}$  is the left inverse operator of $\mathbf{K}_t$ defined by
\begin{equation}
  \mathbf{L}\mathbf{K}_t = \mathbf{I} +  \pi_\circ ,
\label{eq:FPeqfracOU_0040}
\end{equation}
and $\pi_\circ$ is the projection operator given by $\pi_\circ f(t) = f(0)$.
For the three cases in 
(\ref{eq:FPeqfracOU_0020}),
one has
\begin{subequations}
\label{eq:FPeqfracOU_0050}
\begin{numcases}
  {\mathbf{L} = }     I_t                                      \label{eq:FPeqfracOU_0050a}\\
                        I_t^{H+1/2}                               \label{eq:FPeqfracOU_0050b}\\
                        \bigl(1+\lambda^\alpha I_t^\alpha)^{-\gamma}I_t^{\alpha\gamma} . \label{eq:FPeqfracOU_0050c}
\end{numcases}
\end{subequations}
Note that $D_t\pi_\circ = 0$ since the projection operator to the initial point of any function of time
$t$ is a constant.

Recall that the probability of finding the particle in the interval
$x$ and $x + dx$ is given by $P(x,t)dx$, where the probability density
is defined by the average $P(x,t)= \bigl<\rho(x,t)\bigr>_\xi$, and the
corresponding ``equation of motion'' is
\begin{subequations}
  \label{eq:FPeqfracOU_0060}
\begin{eqnarray}
  \partial_t \rho(x,t) & = & - \partial_x \Bigl[D_t\mathbf{L}F(x) + D_t\mathbf{L}\xi\Bigr]\rho(x,t) \nonumber \\
                      & = & -\Bigl[L_\circ + L^\prime\Bigr]\rho(x,t),
\label{eq:FPeqfracOU_0060a}
\end{eqnarray}
where
\begin{equation}
   L_\circ = D_t \mathbf{L}\xi \partial_x, 
   \qquad\qquad
   L^\prime = \partial_x D_t\mathbf{L}F(x) ,
\label{eq:FPeqfracOU_0060b}
\end{equation}
\end{subequations}
Note that the operator $L^\prime$ contains random variable at time $\tau < t$ , 
that means it has to be averaged over random variable $x(\tau)$.

The method used by Reichl 
\cite{Reichl98}
is not applicable to the fractional Langevin
equation 
(\ref{eq:FPeqfracOU_0010}).
Instead, we consider the following alternative Langevin equation
\begin{subequations}
\label{eq:FPeqfracOU_0070}
\begin{equation}
 \mathbf{K}_t x = - \mathbf{K}_t I_t F(x) + \xi ,
\label{eq:FPeqfracOU_0070a}
\end{equation}
or
\begin{equation}
  D_tx = -F(x) + \zeta.
\label{eq:FPeqfracOU_0070b}
\end{equation}
\end{subequations}
(\ref{eq:FPeqfracOU_0070b}) is a Langevin equation with Gaussian fractional noise
$\zeta = D_t\mathbf{L}\xi$.
Note that the force
$F(x)$ 
is local and this is crucial for the use of procedure of 
Reichl \cite{Reichl98}
in deriving the Fokker-Planck equation. 
(\ref{eq:FPeqfracOU_0060})
still holds if
$L^\prime = \partial_x D_t \mathbf{L}F(x)$
is replaced by
$L^\prime = \partial_x F(x)$, 
which has no explicit time dependence. 
Now we define
\begin{equation}
  U(x,t) = e^{-tL^\prime} ,
\label{eq:FPeqfracOU_0080}
\end{equation}
and introduce a new probability density $\vartheta(x,t)$ by
\begin{equation}
  \rho(x,t) = U(x,t)\vartheta(x,t) .
\label{eq:FPeqfracOU_0090}
\end{equation}
From 
(\ref{eq:FPeqfracOU_0060})
and 
(\ref{eq:FPeqfracOU_0090}),
it can be shown that
\begin{equation}
  \partial_t \vartheta(x,t) = - \Theta_x(t)\vartheta(x,t),
\label{eq:FPeqfracOU_0100}
\end{equation}
where
\begin{equation}
  \Theta_x(t) = e^{t\partial_x F(x)}\Bigl[D\mathbf{L}\xi(t)\partial_x\Bigr]e^{-t\partial_x F(x)}.
\label{eq:FPeqfracOU_0110}
\end{equation}
The solution of 
(\ref{eq:FPeqfracOU_0100}) 
is given by
\begin{equation}
  \vartheta(x,t) = e^{-\int_0^t d\tau \Theta_x(\tau)}\vartheta(x,0).
\label{eq:FPeqfracOU_0120}
\end{equation}
By taking the average of $\vartheta(x,t)$ over the white noise $\xi$ gives
\begin{subequations}
  \label{eq:FPeqfracOU_0130}
\begin{eqnarray}
  \bigl<\vartheta(x,t)\bigr> & = & \Bigl<e^{-\int_0^t d\tau \Theta_x(\tau)}\Bigr>\vartheta(x,0)   \label{eq:FPeqfracOU_0130a}\\
                          & = & e^{\frac{1}{2}\bigl<\int_0^t du\int_0^t dv  \Theta_x(u)\Theta_x(v)\bigr>_c}\vartheta(x,0)   \label{eq:FPeqfracOU_0130b} .
\end{eqnarray}
\end{subequations}
The second cumulant of $\Theta_x(t)$ is given by
\begin{eqnarray}\fl
  \Bigl<\int_0^t du\int_0^t dv  \Theta_x(u)\Theta_x(v)\Bigr>_c 
   =  \int_0^t du\int_0^t dv U^{-1}(u)\partial_x U(u) \Bigl<\zeta(u)\zeta(v)\Bigr>_c U^{-1}(v)\partial_x U(v) \nonumber\\
   \qquad \qquad=  \int_0^t du\int_0^t dv U^{-1}(u)\partial_x U(u) C_\zeta(u,v) U^{-1}(v)\partial_x U(v),
  \label{eq:FPeqfracOU_0140}
\end{eqnarray}
where $C_\zeta(u,v)$ is the covariance of the noise $\zeta$.

Now the average of $\vartheta(x,t)$ in (\ref{eq:FPeqfracOU_0130}) can be 
written as
\begin{equation}
\bigl<\vartheta(x,t)\bigr> = e^{\frac{1}{2}\int_0^t du\int_0^t dv U^{-1}(u)\partial_x U(u) C_\zeta(u,v) U^{-1}(v)\partial_x U(v)} \vartheta(x,0).
\label{eq:FPeqfracOU_0150}
\end{equation}
Differentiating 
(\ref{eq:FPeqfracOU_0150})
with respect to time gives
\begin{subequations}
\label{eq:FPeqfracOU_0160}
\begin{eqnarray}
\fl \qquad
  \partial_t\bigl<\vartheta(x,t)\bigr> & = & \left\{
                                \int_0^t d\tau U^{-1}(t)\partial_x U(t) C_\zeta(t,\tau) U^{-1}(\tau)\partial_x U(\tau)
                                \right\} \nonumber \\
\fl                         & & \qquad\quad e^{\frac{1}{2}\int_0^t du\int_0^t dv U^{-1}(u)\partial_x U(u) C_\zeta(u,v) U^{-1}(v)\partial_x U(v)} \vartheta(x,0) 
\label{eq:FPeqfracOU_0160a}\\
\fl                           & = & \left\{
                                \int_0^t d\tau U^{-1}(t)\partial_x U(t) C_\zeta(t,\tau) U^{-1}(\tau)\partial_x U(\tau)
                                \right\}
                                \bigl<\vartheta(x,t)\bigr>.
\label{eq:FPeqfracOU_0160b}
\end{eqnarray}  
\end{subequations}
By taking the average of
(\ref{eq:FPeqfracOU_0090})
with respect to white noise $\xi$ one obtains
\begin{equation}
  P(x,t) = \bigl<\rho(x,t)\bigr> = U(t)\bigl<\vartheta(x,t)\bigr>.
\label{eq:FPeqfracOU_0170}
\end{equation}
Differentiating
(\ref{eq:FPeqfracOU_0170})
with respect to time to get
\begin{eqnarray}\fl
  \ \partial_t P(x,t) & = & - \partial_x F(x)P(x,t) \nonumber \\
\fl                    &   &  + U(t)
                        \left\{
                          \int_0^t d\tau U^{-1}(t)\partial_x U(t) C_\zeta(t,\tau) U^{-1}(\tau)\partial_x U(\tau)
                        \right\}
                        U^{-1}(t)P(x,t) .
\label{eq:FPeqfracOU_0180}
\end{eqnarray}

Let us define operators $\mathbf{S}_x(t)$ and $\Upsilon_x^{\pm}(t)$
which are operators in $x$ and functions of $t$:
\begin{subequations}
\label{eq:FPeqfracOU_0190}
  \begin{eqnarray}
    \Upsilon_x^{+}(t) & = & U^{-1}(t) \partial_x U(t), \label{eq:FPeqfracOU_0190a}\\
    \Upsilon_x^{-}(t) & = & U(t) \partial_x U^{-1}(t), \label{eq:FPeqfracOU_0190b}\\
    \mathbf{S}_x(t) & = & U(t)\left[\int_0^t d\tau \Upsilon_x^{+}(t) C_\zeta (t,\tau)\Upsilon_x^{+}(\tau) \right]U^{-1}(t).
    \label{eq:FPeqfracOU_0190c}
  \end{eqnarray}
\end{subequations}
Using these operators one can write the Fokker-Planck equation corresponds to the Langevin equation 
(\ref{eq:FPeqfracOU_0070})
as
\begin{equation}
  \partial_t P(x,t)  =  - \partial_x F(x)P(x,t) + \mathbf{S}_x(t)P(x,t).
\label{eq:FPeqfracOU_0200}
\end{equation}

For the special case with
$F(x) = 0$, 
the Langevin equation becomes
\begin{equation}
  \mathbf{K}x(t) = \xi(t),
\label{eq:FPeqfracOU_0210}
\end{equation}
or
\begin{equation}
  D_tx(t) = D_t \mathbf{L}\xi(t) =  \zeta(t).
\label{eq:FPeqfracOU_0220}
\end{equation}

Since $L^\prime = \partial_x F(x) = 0$ we have $U(t) = 1$, 
thus in this case one gets
$\Upsilon_x^{+}(t) =  \partial_x$ from (\ref{eq:FPeqfracOU_0190a}),
and
\begin{eqnarray}
  \mathbf{S}_x^\circ(t) & = & \int_0^t d\tau \partial_x C_\zeta(t,\tau) \partial_x 
                         = \frac{1}{2}\dif{}{t}\int_0^t du\int_0^t dv \partial_x C_\zeta(u,v) \partial_x \nonumber \\
                       & = & \frac{1}{2}\dif{}{t}\bigl[C(t,t) - C(t,0) - C(0,t) - C(0,0)\bigr]\partial_x^2 \nonumber \\
                       & = & \frac{1}{2}\dif{\sigma^2(t)}{t}\partial_x^2.
\label{eq:FPeqfracOU_0230}
\end{eqnarray}
Thus the Fokker-Planck equation for the force-free case is given by
\begin{equation}
  \partial_t P(x,t) = \frac{1}{2}\dif{\sigma^2(t)}{t}\partial_x^2 P(x,t).
\label{eq:FPeqfracOU_0240}
\end{equation}
For Brownian motion with variance $\sigma^2(t) = t$,
(\ref{eq:FPeqfracOU_0240})
reduces to the ordinary diffusion equation. 
In the case of fractional Brownian motion with variance
$t^{2H}\Bigl[2H\Gamma\bigl(H - 1/2\bigr)^2\Bigr]^{-1} $
the corresponding Fokker-Planck equation is
\begin{equation}
  \partial_t P(x,t) = \frac{1}{2}\frac{t^{2H-1}}{\Gamma\bigl(H - 1/2\bigr)^2}\partial_x^2 P(x,t),
\label{eq:FPeqfracOU_0241}
\end{equation}
which is in agreement with the result obtained by other authors
\cite{WangLung90,Unal06,HahnKobayashiUmarov11}.

It needs to be pointed out that the effective Fokker-Planck equation does not fully characterize non-Markovian processes such as fractional Brownian motion and fractional Ornstein-Uhlenbeck process. 
For example, both the standard and Riemann-Liouville type fractional Brownian motion, 
and scaled Brownian motion with appropriate scaling factor all have same variance (up to a multiplicative constant), so they all have the same effective Fokker-Planck equation.
Reference 
\cite{LimMuniandy02}
provides a more detailed discussion on this point.

Now consider the case of fractional Ornstein-Uhlenbeck process with two indices.
From its variance given by 
(\ref{eq:BasicfracOU_0020})
one obtains the effective Fokker-Planck equation as
\begin{equation}
  \partial_t P(x,t) = \frac{1}{2}
                      \left[
                        \sum_{q=0}^\infty \bigl(-\lambda^\alpha\bigr)^q
                        \Lambda_qt^{\alpha(2\gamma+q)-2}
                      \right]
                      \partial_x^2 P(x,t),
\label{eq:FPeqfracOU_0250}
\end{equation}
where
\begin{eqnarray}
\fl
  \left[
    \sum_{q=0}^\infty \bigl(-\lambda^\alpha\bigr)^q
    \Lambda_qt^{\alpha(2\gamma+q)-2}
  \right]
   =  \frac{t^{2\alpha\gamma -2}}{\Gamma^2\bigl(\alpha\gamma\bigr)}
        - \lambda^\alpha \frac{2\gamma t^{2\alpha(\gamma+1) -2}}{\Gamma\bigl(\alpha\gamma\bigr)\Gamma\bigl(\alpha(\gamma+1)\bigr)}
        \nonumber \\
\fl \hspace{1.9cm}      + \lambda^{2\alpha} 
            \left[
              \frac{\gamma(\gamma+1)}{\Gamma\bigl(\alpha\gamma\bigr)\Gamma\bigl(\alpha(\gamma+2)}
              + \frac{\gamma^2}{\Gamma\bigl(\alpha(\gamma+1)\bigr)\Gamma\bigl(\alpha(\gamma+1)}
            \right]
            t^{\alpha(2\gamma+2)-2} \ \cdots . 
\label{eq:FPeqfracOU_0260}
\end{eqnarray}

Note that the solution of the Fokker-Planck equation 
(\ref{eq:FPeqfracOU_0240})
subjected to the initial condition
$P(x,0) = \delta(x - x_\circ$
is given by
\begin{equation}
  P(x,t) = \frac{1}{\sqrt{2\pi\sigma^2(t)}}e^{- \frac{|x - x_\circ|^2}{2\sigma^2(t)}}.
\label{eq:FPeqfracOU_0270}
\end{equation}

Note that {(\ref{eq:FPeqfracOU_0260})} contains an infinite number of what appear to be scaled Brownian motion. Fokker-Planck equation with appropriate boundary conditions is widely used in the solving the first
passage time problem. 
In theory it is possible to obtain the first passage time distribution of fractional Ornstein-Uhlenbeck process by using the effective Fokker-Planck equation {(\ref{eq:FPeqfracOU_0260})}, 
in particular its asymptotically large time limit. Such a problem is currently under study.

Finally we remark that in the small time limit, 
(\ref{eq:FPeqfracOU_0250}) reduces to (\ref{eq:FPeqfracOU_0240})
if the Hurst index is taken to be $\alpha\gamma-1/2$. 
This does not come as a surprise since it has been stated earlier that both fractional Brownian motion and Ornstein-Uhlenbeck process satisfy the same local property.


%% file: pathOU2.tex
\section{Path Integral Representation of Fractional Ornstein-Uhlenbeck Process}
\label{sec:pathOU2}
In this section we will obtain
the path integral formulation of the fractional Ornstein-Uhlenbeck process with two indices. For convenience let us denote
\begin{equation}
  \mathbf{K}_t^{\alpha,\gamma}(\lambda) = \bigl(D_t^\alpha + \lambda^\alpha\bigr)^\gamma,
  \label{eq:pathOU2_0010}
\end{equation}
or simply $\mathbf{K}$ since $\alpha$, $\gamma$ and $\lambda$
are fixed throughout this paper.

\subsection{Solution of classical path}
\label{sec:pathOU2classical}
Let us introduce the action for the fractional Ornstein-Uhlenbeck process as follows:
\begin{equation}
  S[x] = \frac{1}{2} \int_0^\beta \bigl[\mathbf{K}\bar{x}(t)\bigr]^2 dt,
\label{eq:pathOU2classical_0010}
\end{equation}
where
\begin{equation}
  \bar{x}(t) = x(t) - x(0).
\label{eq:pathOU2classical_0020}
\end{equation}
The classical solution can be obtained by variational principle,
\begin{equation}
 0 = \delta{S}[x] = \int_0^\beta \left\{
                                   \bigl[D^{\alpha\dagger} + \lambda^\alpha\bigr]^\gamma
                                   \bigl[D^{\alpha} + \lambda^\alpha\bigr]^\gamma
                                   \bar{x}(t)
                                \right\}
                                \delta\bar{x}(t)
                                dt ,
\label{eq:pathOU2classical_0030}
\end{equation}
where the variation at two end points are zero, that is
$\delta\bar{x}(0) = \delta\bar{x}(\beta)=0$.

Denote by $D_t^{\alpha\dagger} = \bigl(D_t^\alpha\bigr)^\dagger$
the adjoint of $D_t^\alpha$ defined by
\begin{equation}
  \int_0^\beta f(t)D_t^\alpha g(t) dt = \int_0^\beta g(t)(D_t^\alpha)^\dagger f(t) dt .
\label{eq:pathOU2classical_0040}
\end{equation}
Let $\lceil{\alpha}\rceil$ be the lowest integer that is greater than or equal to $\alpha$.
We have 
for Riemann-Liouville fractional derivative that
\begin{equation} 
  D_t^\alpha  = D_t^{\lceil{\alpha}\rceil} {_0I_t^{\lceil{\alpha}\rceil -\alpha}} 
  \Longrightarrow 
  \bigl(D_t^\alpha\bigr)^\dagger = {_tI_\beta^{\lceil{\alpha}\rceil -\alpha}}\bigl(-D_t\bigr)^{\lceil{\alpha}\rceil}
                             = {_t^C\!D^\alpha} ,
\label{eq:pathOU2classical_0050x}
\end{equation}
which gives left-fractional derivative of Caputo type ${_t^C\!D^\alpha}$.
On the other hand, the adjoint of the right-fractional derivative of Caputo type
$\bigl({^C\!D_t^\alpha}\bigr)^\dagger$
equals to left-fractional derivative of Riemann-Liouville type 
${_tD^\alpha}$.
Since we consider only the case with zero initial value,
$\bar{x}(0)=0$, 
both
the right-derivative of Caputo and Riemann-Liouville type are equivalent. 
This is, however, not true for the left-derivative.

Using the notation introduced in
(\ref{eq:pathOU2_0010}),
let us consider the adjoint operator
$\mathbf{K}^\dagger$. 
Now the equation of motion can be written as
\begin{equation}
  \mathbf{K}^\dagger \mathbf{K} \bar{x}(t) = 0.
\label{eq:pathOU2classical_0050}
\end{equation}
Consider the derivative of Riemann-Liouville type $D^\alpha$, 
for $0 < \alpha\gamma \leq 1$,
its adjoint $D^{\alpha\dagger}$ is of Caputo-type.
The solution of 
(\ref{eq:pathOU2classical_0050})
can be obtained by first noting
\begin{eqnarray}
   \mathbf{K}^\dagger y(t) & = & \bigl[D_t^{\alpha\dagger} + \lambda^\alpha\bigr]^\gamma y(t)
                           =   \bigl[1 + \lambda^\alpha {_tI^{\alpha}}\bigr]^\gamma {_t^C\!D}^{\alpha\gamma}y(t) \nonumber\\
                         & = &  \bigl[1 + \lambda^\alpha {_tI^{\alpha}}\bigr]^\gamma {_tI^{1-\alpha\gamma}}(-D)y(t)
                           =0,
  \label{eq:pathOU2classical_0060}
\end{eqnarray}
which then gives
\begin{eqnarray}
  y(t) & = & A\sum_{n=0}^\infty \binom{\gamma+n-1}{n} \bigl(-\lambda^\alpha\bigr)^n 
              \frac{(\beta - t)^{\alpha(\gamma+n)-1}}{\Gamma\bigl(\alpha(\gamma+n)\bigr)}  \nonumber \\
       & = & \frac{A}{\Gamma(\gamma)}
             \sum_{n=0}^\infty \frac{\Gamma(\gamma+n)}{n!} \bigl(-\lambda^\alpha\bigr)^n 
              \frac{(\beta - t)^{\alpha(\gamma+n)-1}}{\Gamma\bigl(\alpha(\gamma+n)\bigr)}.
\label{eq:pathOU2classical_0070}
\end{eqnarray}
We can now solve for $\bar{x}$ from the equation 
\begin{subequations}
\label{eq:pathOU2classical_0080}
\begin{equation}
\mathbf{K}\bar{x}(t) = y(t), \label{eq:pathOU2classical_0080a}
\end{equation}
where $\mathbf{K}$
can be reexpressed as 
\begin{equation}
  \mathbf{K} = \bigl[D_t^\alpha + \lambda^\alpha\bigr]^\gamma
            = D_t^{\alpha\gamma}\bigl[1 + \lambda^\alpha I_t^\alpha\bigr]^\gamma 
            = DI_t^{1-\alpha\gamma} \bigl[1 + \lambda^\alpha I_t^\alpha\bigr]^\gamma .
\label{eq:pathOU2classical_0080b}
\end{equation}  
\end{subequations}

It is then straight forward to get the solution by apply 
(\ref{eq:pathOU2classical_0070})
to the right hand side of 
(\ref{eq:pathOU2classical_0080}).
Since the term in summation involved powers of
$\frac{(\beta - t)^{\mu-1}}{\Gamma(\mu)}$, 
it is convenient to consider the following fractional differential equation
\begin{equation}
  D_t^{\alpha\gamma}\bigl[1 + \lambda^\alpha I_t^\alpha\bigr]^\gamma  \bar{x}_\mu(t) = \frac{(\beta - t)^{\mu-1}}{\Gamma(\mu)} ,
\label{eq:pathOU2classical_0090}
\end{equation}
which gives
\begin{equation}
  \bar{x}_\mu(t) = B_\mu + \sum_{m=0}^\infty \binom{\gamma+m-1}{m} \bigl(-\lambda^\alpha\bigr)^m I_t^{\alpha(\gamma+m)}
                          \frac{(\beta - t)^{\mu-1}}{\Gamma(\mu)} .
\label{eq:pathOU2classical_0100}
\end{equation}
By using formula given in 
\cite{GradshteynRyzhik80},
page 317, \#3.197.3, one gets
\begin{equation}
\fl \quad
  I_t^{\alpha(\gamma+m)} \frac{(\beta - t)^{\mu-1}}{\Gamma(\mu)} = \frac{t^{\alpha(\gamma+m)}\beta^{\mu-1}}
                                                                     {\Gamma\bigl(\alpha(\gamma+m)+1\bigr)\Gamma\bigl(\mu\bigr)} 
                                                               {_2F_1}\biggl(1 -\mu, 1, 1 + \alpha(\gamma+m); \frac{t}{\beta}\biggr) .
\label{eq:pathOU2classical_0110}
\end{equation}
Insert
(\ref{eq:pathOU2classical_0110})
into 
(\ref{eq:pathOU2classical_0100})
and then 
(\ref{eq:pathOU2classical_0080})
to obtain the solution, 
which can be written in the following form:
\begin{equation}
  \bar{x}(t) = B + AU(t), 
\label{eq:pathOU2classical_0120}
\end{equation}
where $A$ and $B$ are constants to be determined and
\begin{eqnarray}
\fl \quad
  U(t) & = & \sum_{m=0}^\infty\sum_{n=0}^\infty  \binom{\gamma+m -1}{m}\binom{\gamma+n -1}{n}\bigl(-\lambda^\alpha\bigr)^{m+n} \nonumber \\
\fl       &   & \frac{t^{\alpha(\gamma+m)}\beta^{\alpha(\gamma+n)-1}}
             {\Gamma\bigl(\alpha(\gamma+m)+1\bigr)\Gamma\bigl(\alpha(\gamma+n)\bigr)} 
             {_2F_1}\biggl(1 -\alpha(\gamma + n), 1, 1 + \alpha(\gamma+m); \frac{t}{\beta}\biggr) .
\label{eq:pathOU2classical_0130}
\end{eqnarray}
The values for $A$ and $B$ can be obtained by considering
(\ref{eq:pathOU2classical_0120})
for $t=0$ and $t = \beta$.
It is obvious that $U(0)=0$, which gives $B =0$. 
Consider the value of the hypergeometric function at 
unity, from 
(\cite{GradshteynRyzhik80},
\#9.122.1, p 1008) one gets
\begin{equation}
  {_2F_1}\bigl(\alpha,\nu,\gamma;1\bigr) = \frac{\Gamma(\gamma)\Gamma(\gamma -\alpha-\nu)}{\Gamma(\gamma - \alpha)\Gamma(\gamma - \nu)},
                                         \qquad \bigl[Re\gamma > Re(\alpha+\nu)\bigr] .
\label{eq:pathOU2classical_0140}
\end{equation}
Thus,
\begin{equation}
  U(\beta) = \sum_{q=0}^\infty \bigl(-\lambda^\alpha\bigr)^q \Omega_q \frac{\beta^{\alpha(2\gamma+q)-1}}{\alpha(2\gamma+q) - 1},
\label{eq:pathOU2classical_0150}
\end{equation}
where
\begin{equation}\fl \qquad\qquad
  \Omega_q = \sum_{m+n=q} \binom{\gamma+m-1}{m}\binom{\gamma+n-1}{n}
                         \frac{1}{\Gamma\bigl(\alpha(\gamma+m)\bigr)\Gamma\bigl(\alpha(\gamma+n)\bigr)}.
\label{eq:pathOU2classical_0160}
\end{equation}
Finally, from 
(\ref{eq:pathOU2classical_0120})
for $t=\beta$ gives $A =\frac{\bar{x}(\beta)}{U(\beta)}$,
so the classical path is given by
\begin{eqnarray}\fl \
x_c(t) & = & x_\circ + \frac{x_\beta - x_\circ}{U(\beta)}
                      \sum_{m=0}^\infty \sum_{n=0}^\infty \binom{\gamma+m-1}{m} \binom{\gamma+n-1}{n}\bigl(-\lambda^\alpha\bigr) \nonumber \\
\fl      & & \quad \frac{t^{\alpha(\gamma+m)}\beta^{\alpha(\gamma+n)-1}}{\Gamma\bigl(\alpha(\gamma+m)+1\bigr)\Gamma\bigl(\alpha(\gamma+n)\bigr)}
          {_2F_1}\biggl(1-\alpha(\gamma+n), 1, 1 + \alpha(\gamma+m); \frac{t}{\beta}\biggr).
\label{eq:pathOU2classical_0170}
\end{eqnarray}

\subsection{Propagator}
\label{sec:pathOU2propagator}
Next, we want to evaluate the propagator which is given by
\begin{equation}
  G\bigl(x_\beta,\beta;x_\circ,0\bigr)  = \frac{1}{\mathscr{N}}
                                        \int \mathscr{D}[x]
                                        \delta\bigl(x(\beta) - x_\beta\bigr)
                                        \delta\bigl(x(0) - x_\circ\bigr)
                                        e^{-S[x]} ,
\label{eq:pathOU2propagator_0010}
\end{equation}
where the action is given by
(\ref{eq:pathOU2classical_0010}).

We first consider the transformation
\begin{equation}
  x(t) = x_c(t) + q(t),
\label{eq:pathOU2propagator_0020}  
\end{equation}
where $x_c(\cdot)$ is the classical solution as obtain in 
(\ref{eq:pathOU2classical_0170}), 
and the variable $q(\cdot)$ is the fluctuation
around the classical path satisfying the conditions
$q(0)=q(\beta)=0$.
By substituting 
(\ref{eq:pathOU2propagator_0010})
into 
(\ref{eq:pathOU2classical_0010})
gives the action as
\begin{eqnarray}\fl\qquad
  S[x] & = & \frac{1}{2} \int_0^\beta \bigl[\mathbf{K}\bar{x}_c(t)\bigr]^2 dt
       + \int_0^\beta \bigl[\mathbf{K}\bar{x}_c(t)\bigr]\bigl[\mathbf{K}q(t)\bigr] dt,
       + \frac{1}{2} \int_0^\beta \bigl[\mathbf{K}q(t)\bigr]^2 dt 
\label{eq:pathOU2propagator_0030}  \\
\fl    & = & S[x_c]+S[q].
\label{eq:pathOU2propagator_0040}  
\end{eqnarray}
The cross terms disappear since $\mathbf{K}^\dagger\mathbf{K}\bar{x}_c = 0$
(see (\ref{eq:pathOU2classical_0050})).
Now substituting 
(\ref{eq:pathOU2propagator_0040})
into 
(\ref{eq:pathOU2propagator_0010})
to get
\begin{equation}
  G\bigl(x_\beta,\beta;x_\circ,0\bigr) = e^{-S[x_c]} G\bigl(x_\circ,\beta;x_\circ,0\bigr) ,
\label{eq:pathOU2propagator_0050}
\end{equation}
where
$G\bigl(x_\circ,\beta;x_\circ,0\bigr)$
is the loop part, i.e. the propagator that begin and end at the same
point $x(\beta)=x(0)=x_\circ$.

Recall that $y_c(t) = \mathbf{K}\bar{x}_c(t)$,
and it is given by
(\ref{eq:pathOU2classical_0070})
with the constant $A$ previously
determined. To be more precise, one can write
\begin{subequations}
\label{eq:pathOU2propagator_0060}
  \begin{equation}
    y_c(t)      =  \bigl(x_\beta - x_\circ\bigr)W(\beta-t)  ,
    \label{eq:pathOU2propagator_0060a}
  \end{equation}
  \begin{equation}
    W(\beta-t)   =  \frac{1}{U(\beta)}
                     \sum_{n=0}^\infty \binom{\gamma+n-1}{n}\bigl(-\lambda^\alpha\bigr)^n 
                     \frac{(\beta-t)^{\alpha(\gamma+n)-1}}{\Gamma\bigl(\alpha(\gamma+n)\bigr)}.
  \label{eq:pathOU2propagator_0060b}
  \end{equation}
\end{subequations}
Thus, the action becomes
\begin{equation}
  S[x_c] = \frac{|x_\beta - x_\circ|^2}{2}\int_0^\beta W^2(\beta-t).
\label{eq:pathOU2propagator_0070}
\end{equation}
The integral term in the above equation can be computed:
\begin{eqnarray}\fl \quad
  \int_0^\beta W^2(\beta-t) & = & \frac{1}{\bigl[U(\beta)\bigr]^2}
                                \sum_{m=0}^\infty\sum_{n=0}^\infty
                                \binom{\gamma+m-1}{m}\binom{\gamma+n-1}{n}
                                \bigl(-\lambda^\alpha\bigr)^{m+n} \nonumber \\
\fl                           & & \qquad \frac{1}{\Gamma\bigl(\alpha(\gamma+m)\bigr)\Gamma\bigl(\alpha(\gamma+n)\bigr)}
                                \int_0^\beta dt (\beta-t)^{\alpha(2\gamma + m+n)-2} .
\label{eq:pathOU2propagator_0080}
\end{eqnarray}
For $\alpha\gamma > 1/2$,
the summation term exactly equals to $U(\beta)$,
which gives
\begin{equation}
  \int_0^\beta W^2(\beta-t) = \frac{1}{U(\beta)} .
\label{eq:pathOU2propagator_0090}
\end{equation}
Thus, one obtains
\begin{equation}
  G\bigl(x_\beta,\beta;x_\circ,0\bigr) = e^{-\frac{|x_\beta-x_\circ|^2}{2U(\beta)}} G\bigl(x_\circ,\beta;x_\circ,0\bigr)  .
\label{eq:pathOU2propagator_0100}
\end{equation}
and
\begin{equation}\fl\qquad
  1 = \int_{-\infty}^\infty dx_\beta e^{-\frac{|x_\beta-x_\circ|^2}{2U(\beta)}} G\bigl(x_\circ,\beta;x_\circ,0\bigr) 
    = \sqrt{2\pi{U(\beta)}} G\bigl(x_\circ,\beta;x_\circ,0\bigr) .
\label{eq:pathOU2propagator_0110}
\end{equation}
Finally, the propagator is obtained as
\begin{equation}
   G\bigl(x_\beta,\beta;x_\circ,0\bigr) = \frac{1}{ \sqrt{2\pi{U(\beta)}}}e^{-\frac{|x_\beta-x_\circ|^2}{2U(\beta)}}.
\label{eq:pathOU2propagator_0120}
\end{equation}

Here we note that the variance
\begin{equation}\fl\quad \
   \mathscr{E}\Bigl[\bigl((x(\beta) - x(0)\bigr)^2 \Bigl| x(0) = x_\circ\Bigr.\Bigr]
                    = \int_{-\infty}^\infty dx_\beta \bigl|(x_\beta - x_\circ\bigr|^2  G\bigl(x_\beta,\beta;x_\circ,0\bigr)
                    = U(\beta).
\label{eq:pathOU2propagator_0130}
\end{equation}

\subsection{Partition Function}
\label{sec:pathOU2partition}
The partition function, which is defined as the trace of the propagator, is given by
\begin{equation}
  Z(\beta) = \int_V dx G\bigl(x,\beta;x,0\bigr) = \frac{V}{\sqrt{2\pi{U(\beta)}}} ,
\label{eq:pathOU2partition_0010}
\end{equation}
where $V$ is the volume enclosing the particle.

\subsection{Generating function}
\label{sec:pathOU2generating}
The generating function is defined as logarithm of the partition function with source term
$\bigl<h,\bar{x}\bigr>$:
\begin{equation}
  Z(\beta,h) = Z(\beta) e^{\frac{1}{2}\bigl<h,\mathbf{R}\mathbf{R}^\dagger h\bigr>}.
\label{eq::pathOU2generating_0010}
\end{equation}
where $\mathbf{R}$ is the right inverse of $\mathbf{K}$, i.e. $\mathbf{KR} = I$ an indentity.

Now consider the free energy
\begin{equation}
  F(\beta,h) = - \frac{1}{\beta}\log{Z(\beta,h)} = F(\beta,0) - \frac{1}{2\beta}\bigl<h,\mathbf{R}\mathbf{R}^\dagger h\bigr>.
\label{eq::pathOU2generating_0020}
\end{equation}
We then have
\begin{equation}
  \frac{\delta{F(\beta,h)}}{\delta{h(t)}} = - \frac{1}{\beta Z(\beta,h)}\frac{\delta{Z(\beta,h)}}{\delta{h(t)}}
                                          = \frac{1}{\beta}\mathscr{E}_h\bigl[\bar{x}(t)\bigr] .
\label{eq::pathOU2generating_0030}
\end{equation}
The mean of $\bar{x}$ is given by
\begin{equation}
  M(t) = \mathscr{E}_\circ\bigl[\bar{x}(t)\bigr] = \beta \lim_{h\to 0}\frac{\delta{F(\beta,h)}}{\delta{h(t)}}
                                                = \lim_{h\to 0}\mathbf{R}\mathbf{R}^\dagger h(t) = 0.
\label{eq::pathOU2generating_0040}
\end{equation}
Here, $\mathscr{E}_h(\cdot)$ and $\mathscr{E}_\circ(\cdot)$ are the expectations with source $h$ and zero, respectively.
Similarly, 
the covariance can be evaluated as
\begin{eqnarray}
  C(s,t) & = & \beta \lim_{h\to 0}\frac{\delta^2}{\delta{h(s)}\delta{h(t)}}F(\beta,h) 
         =\mathbf{R}\mathbf{R}^\dagger \delta(t-s) .
\label{eq::pathOU2generating_0050}
\end{eqnarray}
This is the same as the result obtained by using the Langevin equation (with the
probability distribution of the random variable restricted to the finite region $[0,\beta]$),
(see \ref{sec:detail_cov}).


%% file: conclude.tex
\section{Concluding remarks}
\label{sec:conclude}
We have obtained the Fokker-Planck equation for the fractional 
Ornstein-Uhlenbeck process with two indices. 
This effective Fokker-Planck equation is more
complicated than that for fractional Brownian motion due to the complex structure of the
variance of the fractional Ornstein-Uhlenbeck process. The path integral representation of
fractional Ornstein-Uhlenbeck process with two indices of Riemann-Liouville type and various relevant physical quantities are derived.

Physical applications of fractional Ornstein-Uhlenbeck process with two indices to the Von Karman wind speed
spectrum 
\cite{LimLiTeo08}
and Casimir energy for fractional quantum field have been considered 
\cite{LimTeo09,LimTeo11}.
Note that in the latter application one makes use of the fact
that fractional Ornstein-Uhlenbeck process with two indices can be regarded
as one-dimensional fractional Klein-Gordon scalar massive Euclidean field. 
In view of the recent interest of quantum field theories with multifractal spacetime structure
in the sense that spacetime is of integer dimension 4 at large scales, and it is two dimensional
in the small scales
\cite{Modesto09,Ambjorn10,Calcagni10,Calcagni13,ReuterSaueressig12,Calcagni12},
it is hoped that the path integral formulation given here
may have relevance in some of these theories.


%% file: FPeqPotential.tex
\section{Fokker-Planck equation with some simple potentials}
\label{sec:FPeqPotential}
Here we derive the values of $\Upsilon$, hence $\mathbf{S}_x(t)$, 
which allows one to obtain the associated Fokker-Planck equation as given by
(\ref{eq:FPeqfracOU_0200}).
From (\ref{eq:FPeqfracOU_0080}) with $L^\prime = \partial_xF(x)$
 and (\ref{eq:FPeqfracOU_0190a}), we write explicitly
\begin{equation}
  \Upsilon^{\pm}_x(t) = e^{{\pm}t\partial_xF(x)} \partial_x e^{{\mp} t\partial_xF(x)}.
\label{eq:FPeqPotential_0010}
\end{equation}
Differentiate (\ref{eq:FPeqPotential_0010}) with respect to time gives
\begin{eqnarray}
  \partial_t \Upsilon^{\pm}_x(t) & = &{\mp}e^{{\pm} t\partial_xF(x)} \bigl[\partial_x,\partial_xF(x)\bigr] e^{{\mp}t\partial_xF(x)} \nonumber \\
                           & = & {\mp}e^{{\pm}t\partial_xF(x)} \partial_x\bigl[\partial_x,F(x)\bigr] e^{{\mp}t\partial_xF(x)} . 
\label{eq:FPeqPotential_0020}
\end{eqnarray}

The cases to be considered are given in the Table \ref{tab:potential}.

\begin{table}[H]
  \centering
\begin{tabular}{|l|c|c|c|c|}\hline 
          & $V(x)$ & $F(x)$ & $U(t)$  & $\bigl[\partial_x,F(x)\bigr]$\rule{0pt}{12pt} \\[2pt] \hline
  Free    & $c$    & $0$    & 1       & $0$  \\ \hline
  Linear  & $gx$   & $- g$  & $e^{tg\partial_x}$    & 0  \\ \hline
  Hamonic & $\frac{1}{2}\omega x^2$ & $-\omega x$ & $e^{t\omega\partial_x x}$ & $-\omega $ \\ \hline
\end{tabular}  
  \caption{Some simple potentials}
  \label{tab:potential}
\end{table}

For the free and linear cases, (\ref{eq:FPeqPotential_0010}) becomes
  \begin{eqnarray}
    \partial_t \Upsilon^{+}_x(t) & = & 0 , \label{eq:FPeqPotential_0030a}\\
    \Upsilon^{+}_x(t) & = & \Upsilon^{+}_x(0) =  \partial_x . \label{eq:FPeqPotential_0030b}
  \end{eqnarray}
Thus the corresponding operator $\mathbf{S}_x(t)$ is 
\begin{equation}
  \mathbf{S}_x(t) = e^{tg\partial_x}\frac{1}{2}\frac{d\sigma^2(t)}{dt}\partial_x^2e^{-tg\partial_x}     
                  = \frac{1}{2}\frac{d\sigma^2(t)}{dt}\partial_x^2,
\end{equation}
and thus it is the same as free case (\ref{eq:FPeqfracOU_0230}),
the Fokker-Planck equation for the linear case is given by
\begin{equation}
  \partial_t P(x,t) = -g\partial_x P(x,t)
                      + \frac{1}{2}\frac{d\sigma^2(t)}{dt}\partial_x^2 P(x,t).
\label{eq:FPeqPotential_0040}
\end{equation}

Finally, for the case of harmonic potential, (\ref{eq:FPeqPotential_0010}) can be written as
\begin{eqnarray}
  \partial_t \Upsilon^{\pm}_x(t) & = & {\pm}\omega \Upsilon^{\pm}_x(t),  \\
             \Upsilon^{\pm}_x(t) & = & \Upsilon^{\pm}_x(0)e^{\pm\omega{t}} = \partial_x e^{\pm\omega{t}} .
\label{eq:FPeqPotential_0050}
\end{eqnarray}
Therefore one gets
\begin{equation}
 \mathbf{S}_x(t) = \Bigl[\int_0^t d\tau C_\zeta(t,\tau)e^{(t+\tau)\omega}\Bigr]U(t)\partial_x^2U^{-1}(t).                 
\label{eq:FPeqPotential_0060}
\end{equation}
Remark that 
\begin{equation}
  U(t)\partial_x^2 U^{-1}(t) = \bigr[\Upsilon^{-}_x(t)\bigl]^2 = \partial_x^2 e^{-2\omega{t}}.
\end{equation}
Thus we have
\begin{equation}
 \mathbf{S}_x(t) = \Bigl[\int_0^t d\tau C_\zeta(t,\tau)e^{-\omega(t-\tau)}\Bigr]\partial_x^2 .
\end{equation}
Thus the Fokker-Plack equation for this case is
\begin{equation}
  \partial_t P(x,t) = \omega\partial_x xP(x,t)
                      + \Bigl[\int_0^t d\tau C_\zeta(t,\tau)e^{-\omega(t-\tau)}\Bigr]\partial_x^2 P(x,t).
\label{eq:FPeqPotential_0070}
\end{equation}


%% file: pathOU2detail.tex
\section{Evaluations of generating function and covariance}
\label{sec:details}

\subsection{Generating function}
\label{sec:detail_gen}
The following argument is used to obtain (\ref{eq::pathOU2generating_0010}).
Add the source $\bigl<h,x\bigr>$ to the action and use
the same procedure as in subsection {\ref{sec:pathOU2classical}} and instead of 
(\ref{eq:pathOU2classical_0050}) we get 
\begin{equation}
  \mathbf{K}^\dagger\mathbf{K}x(t) + h(t) = 0.
\label{eq:detail_gen_0010}
\end{equation}
Thus, 
\begin{equation}
  x(t) = x^\circ(t) - \mathbf{R}\mathbf{R}^\dagger h(t) ,
\label{eq:detail_gen_0020}
\end{equation}
where $x^\circ$ is the homogeneous solution, i.e. when $h = 0$.
The similar procedure is repeated till we get the propagator
with source $h$:
\begin{equation}
  G\bigl[x_\beta,\beta;x_\circ,0;h\bigr] = G\bigl[x_\beta,\beta;x_\circ,0;0\bigr]e^{\frac{1}{2}\Bigl<h,\mathbf{R}\mathbf{R}^\dagger h\Bigr> } ,
\label{eq:detail_gen_0030}
\end{equation}
where the propagator $G\bigl[x_\beta,\beta;x_\circ,0;0\bigr]$ is the same as 
(\ref{eq:pathOU2propagator_0120}).
By taking the trace of $\int_V dx G\bigl[x,\beta;x,0;h\bigr]$  the same as in (\ref{eq:pathOU2partition_0010}), we get
(\ref{eq::pathOU2generating_0010}).

\subsection{Covariance}
\label{sec:detail_cov}
Let express the last term of (\ref{eq::pathOU2generating_0050}) in details:
\begin{eqnarray}\fl\quad \;
  \mathbf{R}\mathbf{R}^\dagger \delta(t-s) & = & \int_0^t du R(t,u)\int_u^\beta dv R(v,u)\delta(v -s)  \nonumber\\
                                         & = &  \int_0^\beta du \Theta(t-u)R(t,u)\int_0^\beta dv \Theta(v-u)R(v,u)\delta(v -s)  \nonumber\\
                                         & = &  \int_0^\beta du \Theta(t-u)R(t,u)\Theta(s-u)R(s,u) \nonumber \\
                                         & = &  \int_0^{\min(t,s)} duR(t,u)R(s,u) ,
\end{eqnarray}
which is the same as that obtained from the Langevin equation:
\begin{eqnarray}
  C(t,s) & = &\Bigl<\mathbf{R}\xi(t)\mathbf{R}\xi(s)\Bigr>_\xi  \nonumber\\
         & = & \int_0^t du R(t,u)\int_0^s dv R(s,v) \bigl<\xi(u)\xi(v)\bigr>_\xi  \nonumber\\
         & = & \int_0^t du R(t,u)\int_0^s dv R(s,v) \delta(u -v).
\end{eqnarray}


%% file: CaputoRL.tex
\section{Caputo type equation}
\label{sec:CaputoRL}
In this appendix we discuss the solution of equation of motion (\ref{eq:pathOU2classical_0060}) of Caputo type.
It is straightforward to show that {(\ref{eq:pathOU2classical_0070})} is the solution of Riemann-Liouville type
\begin{equation}
   {_tD^{\alpha\gamma}}\bigl[1 + \lambda^\alpha {_tI^{\alpha}}\bigr]^\gamma y(t) = 0.
\label{eq:CaputoRL_0010}
\end{equation}
Now, we insert identity $(-D)({_tI})=\mathbf{I}$ into l.h.s.\@ of (\ref{eq:pathOU2classical_0060}) to give
\begin{eqnarray}\fl
(-D)({_tI})\bigl[1 + \lambda^\alpha {_tI^{\alpha}}\bigr]^\gamma {_tI^{1-\alpha\gamma}}(-D)y^C(t) 
        & = & (-D)\bigl[1 + \lambda^\alpha {_tI^{\alpha}}\bigr]^\gamma {_tI^{1-\alpha\gamma}}{_tI}(-D) y^C(t) \nonumber \\
        & = & (-D)\bigl[1 + \lambda^\alpha {_tI^{\alpha}}\bigr]^\gamma {_tI^{1-\alpha\gamma}}\bigl[y^C(t)-y_\beta^C \bigr]  \nonumber\\
        & = & (-D){_tI^{1-\alpha\gamma}}\bigl[1 + \lambda^\alpha {_tI^{\alpha}}\bigr]^\gamma \bigl[y^C(t)-y_\beta^C \bigr]  \nonumber\\
        & = & {_tD^{\alpha\gamma}}\bigl[1 + \lambda^\alpha {_tI^{\alpha}}\bigr]^\gamma \bigl[y^C(t)-y_\beta^C \bigr] ,
\label{eq:CaputoRL_0020}
\end{eqnarray}
which is clear that 
\begin{equation}
  y^C(t)  = y_\beta^C + y(t).
\label{eq:CaputoRL_0030}
\end{equation}
The undetermined constant $y_\beta^C$ can be chosen to be zero,
thus minimizes the action (\ref{eq:pathOU2classical_0010}).
This verify that the solution of Riemann-Lioville type is also a soltution of Caputo type.


%% file: main.bbl
\providecommand{\newblock}{}
\begin{thebibliography}{10}
\expandafter\ifx\csname url\endcsname\relax
  \def\url#1{{\tt #1}}\fi
\expandafter\ifx\csname urlprefix\endcsname\relax\def\urlprefix{URL }\fi
\providecommand{\eprint}[2][]{\url{#2}}

\bibitem{Biagini08}
Biagini F, Hu Y, \O{}ksendal B and Zhang T 2008 {\em Stochastic Calculus for
  Fractional {B}rownian Motion and Applications\/} (Springer, New York)

\bibitem{Mishura08}
Mishura Y 2008 {\em Stochastic Calculus for Fractional Brownian Motion and
  Related Processes\/} (Springer, New York)

\bibitem{Nourdin12}
Nourdin I 2012 {\em Selected Aspects of Fractional {B}rownian Motion\/}
  (Springer, New York)

\bibitem{Kalages08}
Klages R, Radons G and Sokolov I (eds) 2008 {\em Anomalous Transport:
  Foundations and Applications\/} (Wiley-VCH, Weinheim, Germany)

\bibitem{Cheridito03}
Cheridito P, Kawaguchi H and Maejima M 2003 {\em Electronic J. Probability\/}
  {\bf 8} {paper no. 3,} 1--14

\bibitem{LimMuniandy03}
Lim S~C and Muniandy S~V 2003 {\em J. Phys. A\/} {\bf 36} 3961--3982

\bibitem{Magdziarz08}
Magdziarz M 2008 {\em Physica A\/} {\bf 387} 123--133

\bibitem{YanLuXu08}
Yan L, Lu Y and Xu Z 2008 {\em J. Phys.A\/} {\bf 41} 145007 (17 pages)

\bibitem{Chaichian01}
Chaichian M and Demichev A 2001 {\em Path Integrals in Physics Volume {I}:
  Stochastic Processes and Quantum Mechanics\/} (IOP Publishing, Bristol)

\bibitem{Kleinert09}
Kleinert H 2009 {\em Path Integral, in Quantum Mechanics, Statistics, Polymer
  Physics and Financial Markets\/} 5th ed (World Scientific, Singapore)

\bibitem{Wio13a}
Wio H~S 2013 {\em Path Integrals for Stochastic Processes: An Introduction\/}
  (World Scientific, Singapore)

\bibitem{Laskin00}
Laskin N 2000 {\em Phys. Lett. A\/} {\bf 268} 298--305

\bibitem{TarasovZaslavsky08}
Tarasov V~E and Zaslavsky G~M 2008 {\em Communications in Nonlinear Science and
  Numerical Simulation\/} {\bf 13} 248--258

\bibitem{CalcagniNardelliScalisi12}
Calcagni G, Nardelli G and Scalisi M 2012 {\em J. Math. Phys.\/} {\bf 53}
  102110

\bibitem{KleinertZatloukal13}
Kleinert H and Zatloukal V 2013 {\em Phys. Rev. E\/} {\bf 88} 052106

\bibitem{Kleinert12}
Kleinert H 2012 {\em Europhys. Lett.\/} {\bf 100} 10001

\bibitem{Ferrante13}
Ferrante D~D, Guralnik G~S, Guralnik Z and Pehlevan C 2013 {\em arXiv:\/}
  1301.4233v2 [hep--th] 3 Mar 2013

\bibitem{MagnenUnterberger11}
Magnen J and Unterberger J 2011 {\em Ann. H. Poincar\'{e}\/} {\bf 12}
  1199--1226

\bibitem{MagnenUnterberger12}
Magnen J and Unterberger J 2012 {\em Ann. H. Poincar\'{e}\/} {\bf 13} 209--270

\bibitem{Modesto09}
Modesto L 2009 {\em Class. Quant. Grav.\/} {\bf 26} 242002

\bibitem{Ambjorn10}
Ambj\o{}rn J, G\"{o}rlich A, Jurkiewicz J and Loll R 2010 {\em Phys. Lett. B\/}
  {\bf 690} 420--426

\bibitem{Calcagni10}
Calcagni G 2010 {\em Phys. Rev. Lett.\/} {\bf 104} 251301

\bibitem{Calcagni13}
Calcagni G 2013 {\em Phys. Rev. D\/} {\bf 88} 065005

\bibitem{ReuterSaueressig12}
Reuter M and Saueressig F 2012 {\em New J. Phys.\/} {\bf 14} 055022

\bibitem{Calcagni12}
Calcagni G 2012 {\em JHEP\/} {\bf 1201} 065

\bibitem{Sebastian95}
Sebastian K 1995 {\em Journal of Physics A: Mathematical and General\/} {\bf
  28} 4305--4311

\bibitem{CherayilBiswas93}
Cherayil B~J and Biswas P 1993 {\em J. Chem. Phys.\/} {\bf 99} 9230--9236

\bibitem{ChakravartiSebastian97}
Chakravarti N and Sebastian K~L 1997 {\em Chem. Phys. Lett.\/} {\bf 267} 9--13

\bibitem{CalvoSanchez08}
Calvo I and S\'{a}nchez R 2008 {\em J. Phys. A: Math. Theor.\/} {\bf 41} 282002

\bibitem{CalvoSanchezCarreras09}
Calvo I, S\'{a}nchez R and Carreras B 2009 {\em J. Phys. A: Math. Theor.\/}
  {\bf 42} 055003

\bibitem{JanakiramanSebastian12}
Janakiraman D and Sebastian K 2012 {\em Phys. Rev. E\/} {\bf 86} 061105

\bibitem{Wio13b}
Wio H 2013 {\em J. Phys. A: Math. Theor.\/} {\bf 46} 115005

\bibitem{EabLim06}
Eab C~H and Lim S~C 2006 {\em Physica A\/} {\bf 371} 303--316

\bibitem{FriedrichEule11}
Friedrich R and Eule S 2011 {\em arXiv:1110.5771v1 [cond-mat.stat-mech]\/}

\bibitem{EuleFriedrich07}
Eule S and Friedrich R 2007 Towards a path-integral formulation of continuous
  time random walks {\em Path Integrals - New Trends and Perspectives\/} (World
  Sci. Publ., Singapore) pp 581--584

\bibitem{PeltierVehel95}
Peltier R and Vehel J~L 1995 Multifractional brownian motion : definition and
  preliminary results. Tech. rep. Rapport de recherche de l'INRIA, 2645

\bibitem{BenassiJaffardRoux97}
Benassi A, Jaffard S and Roux D 1997 {\em Rev. Mat. Iberoam\/} {\bf 13} 19--81

\bibitem{LimTeo07}
Lim S and Teo L 2007 {\em J. Phys. A: Math. Gen.\/} {\bf 40} 6035--6060

\bibitem{LimLiTeo08}
Lim S~C, Li M and Teo L~P 2008 {\em Phys. Lett. A\/} {\bf 372} 6309--6320

\bibitem{LimTeo09}
Lim S~C and Teo L~P 2009 {\em J. Phys. A: Math. Theor.\/} {\bf 42} 065208
  (34pp).

\bibitem{LimTeo11}
Lim S and Teo L 2011 Casimir effect associated with fractional klein-gordon
  field {\em Fractional Dynamics, Recent Advances\/} ed Klafter Y, Lim S~C and
  Metzler R ((World Scientific, Singapore)) pp 483--506

\bibitem{LimEab06}
Lim S~C and Eab C~H 2006 {\em Phys. Lett. A\/} {\bf 355} 87--96

\bibitem{Ezawa00}
Ezawa H 2000 {\em Acta Appl. Math.\/} {\bf 63} 119--135

\bibitem{BelBarkai05}
Bel G and Barkai E 2005 {\em Phys. Rev. Lett.\/} {\bf 94} 240602

\bibitem{DengBarkai09}
Deng W and Barkai E 2009 {\em Phys. Rev. E\/} {\bf 79}(1) 011112

\bibitem{SamkoKibasMarichev93}
Samko S~G, Kibas A~A and Marichev O~I 1993 {\em Fractional Integral and
  Derivatives: Theory and Applications\/} (Amsterdam: Gordonand Breach)

\bibitem{Samko02}
Samko S~G 2002 {\em Hypersingular Integrals and Their Applications\/} ({\em
  Series: Analytical Methods and Special Functions\/} vol~5) (Taylor \&
  Francis)

\bibitem{Lim01}
Lim S~C 2001 {\em J. Phys. A: Math. Gen.\/} {\bf 34} 1301--1310

\bibitem{MagdziarzWeron07}
Magdziarz M and Weron A 2007 {\em Studia Mathematica\/} {\bf 181(1)} 47--60

\bibitem{Magdziarz09}
Magdziarz M 2009 {\em Stochastic Processes and their Applications\/} {\bf 119}
  3416 -- 3434 ISSN 0304-4149

\bibitem{MagdziarzWeron11}
Magdziarz M and Weron A 2011 {\em Annals of Physics\/} {\bf 326} 2431--2443

\bibitem{Khinchin48}
Khinchin A 1948 {\em Mathematical Foundations of Statistical Mechanics\/}
  ((Dover, New York).)

\bibitem{Reichl98}
Reichl L~E 1998 {\em A Modern course in Statistical Physics\/} 2nd ed (John
  Wiley \& Sons)

\bibitem{WangLung90}
Wang K and Lung C 1990 {\em Physics Letters A\/} {\bf 151} 119--121

\bibitem{Unal06}
\"{U}nal G 2007 {F}okker-{P}lanck-{K}olmogorov equation for f{B}m: derivation
  and analytical solutions {\em Mathematical Physics, Proceedings of the 12th
  Regional Conference, Islamabad, Pakistan, 27 March {-} 1 April 2006\/} (World
  Sci. Publ., Singapore) pp 53--60

\bibitem{HahnKobayashiUmarov11}
Hahn M~G, Kobayashi K and Umarov S 2011 {\em Proc. Amer. Math. Soc.\/} {\bf
  139} 691--705

\bibitem{LimMuniandy02}
Lim S~C and Muniandy S~V 2002 {\em Phys. Rev. E\/} {\bf 66}(2) 021114

\bibitem{GradshteynRyzhik80}
Gradshteyn I~S and Ryzhik I~M 1980 {\em Table of Integrals, Series, and
  Products\/} 4th ed (Academic Press, Inc.)

\end{thebibliography}
